# Development of a skateboarding trick classifier using accelerometry and machine learning

Nicholas Kluge Corrêa[1]*, Júlio César Marques de Lima[1], Thais Russomano[2], Marlise Araujo dos Santos[3]

[1] Aerospace Engineering Laboratory, Microgravity Center, Faculty of Electrical Engineering, Pontifical Catholic University of Rio Grande do Sul, Porto Alegre, RS, Brazil.
[2] Center of Human and Aerospace Physiological Sciences, Faculty of Life Sciences & Medicine, King's College, London, England.
[3] Joan Vernikos Aerospace Pharmacy Laboratory, Microgravity Center, Faculty of Electrical Engineering, Pontifical Catholic University of Rio Grande do Sul, Porto Alegre, Brazil.

**Abstract** **Introduction:** Skateboarding is one of the most popular cultures in Brazil, with more than 8.5 million skateboarders. Nowadays, the discipline of street skating has gained recognition among other more classical sports and awaits its debut at the Tokyo 2020 Summer Olympic Games. This study aimed to explore the state-of-the-art for inertial measurement unit (IMU) use in skateboarding trick detection, and to develop new classification methods using supervised machine learning and artificial neural networks (ANN). **Methods:** State-of-the-art knowledge regarding motion detection in skateboarding was used to generate 543 artificial acceleration signals through signal modeling, corresponding to 181 flat ground tricks divided into five classes (NOLLIE, NSHOV, FLIP, SHOV, OLLIE). The classifier consisted of a multilayer feed-forward neural network created with three layers and a supervised learning algorithm (backpropagation). **Results:** The use of ANNs trained specifically for each measured axis of acceleration resulted in error percentages inferior to 0.05%, with a computational efficiency that makes real-time application possible. **Conclusion:** Machine learning can be a useful technique for classifying skateboarding flat ground tricks, assuming that the classifiers are properly constructed and trained, and the acceleration signals are preprocessed correctly.

# Introduction

Skateboarding over the last decade has become a popular sport in Brazil. A 2009 study conducted by the Datafolha Research Institute aimed to measure the size of the skateboarding community in the Brazilian population, revealing that 3.8 million individuals were practicing the sport at that time. The survey was repeated in 2015 and showed a 123% growth in participants, with 8.5 million skateboarders (Confederação…, 2016).

The International Olympic Committee (IOC) made an important worldwide announcement regarding skateboarding on the 3rd August 2016, agreeing to the inclusion of the sport of skateboarding in the Tokyo 2020 Olympic Games. The Tokyo 2020 Skateboarding Commission (TSC), together with the IOC, will oversee the insertion and regulation of the sport of skateboarding in the Olympics (Tokyo…, 2016).

Modern skateboarding (Street) uses the urban architecture (banks, walls, steps, handrails) as obstacles, with the skateboarder making use of a variety of different types of tricks to overcome the obstacles (Candotti et al., 2012). Research has shown the practice of skateboarding in urban regions to present a greater risk of injuries and accidents (Everett, 2002), where the most affected are children and untrained beginners (Department…, 2013; Fountain and Meyers, 1996; Forsman and Eriksson, 2001). Given the growing popularity of this sport and its intrinsic risk factor, alternative ways to assist the development of this discipline should be pursued, and the use of computer-assisted assessment (CAA) may be a promising tool.

Exertion Games, also called exergames, are games where the application of physical effort by the user is necessary (Muller et al., 2008). Studies have already proven that interactive simulations can lead to an improvement in sports performance, such as with skating and ice speed skating (Stienstra et al., 2011), rowing (Broker and Crawley, 2001), and even swimming (Marc et al., 2009).

Skateboarding-based exergames currently have no realistic interactive interface; the controls generally used by these games do not allow true movement or control of the skateboard. Nevertheless, advances in the development of micro-electromechanical sensors and inertial measurement units (IMUs) can assist in developing tools giving more realism (Skog and Händel, 2006).

The Tilt 'n' Roll was a prototype game developed by Anlauff et al. (2010), whereby a conventional skateboard was equipped with IMU sensors (3D accelerometer and gyroscope) that allowed the accelerations imposed on the skateboard to be monitored, and the differentiation of three tricks; Ollie, Frontside Ollie, and Bail (incorrect attempt/fall). A linear discriminant analysis (LDA) was used for trick classification, with 0.5 s windows to detect events in the IMU signal, sampled at 70 samples/s. The LDA classifier was modeled through a database composed of 20 tricks collected for each class. The classification method showed a 90% true positive classification, and real-time application was possible in modern smartphones (Anlauff et al., 2010).

More recently, IMU sensors (3D accelerometer and gyroscope) combined with a trick detection algorithm have been developed, in which four different classification methods (Naive Bayes, Partial Decision Tree, Support Vector Machine, K-Nearest Neighbor) were compared to distinguish between six different tricks; Ollie, Nollie, Kickflip, Heelflip, Pop Shove-it and 360-Kickflip (Groh et al., 2015). With a scale adjusted to 16g and 2000°/s, and signals sampled at 200 sample/s, the detection algorithm was able to point correctly to 323 of the 343 signals produced, resulting in a sensitivity of 94.2%. The Naive Bayes and Support Vector Machine methods achieved the best results, with 97.8% of true positives. However, both these classification methods developed in the study produced a considerable computational effort, given the great number of extracted features for each trick detected (Groh et al., 2015).

The aforementioned studies used similar methods of data acquisition and detection, and both used linear regression techniques to classify the sample space. Furthermore, both studies pointed to a need for more efficient and faster classification algorithms so that larger groups of tricks can be detectable, especially considering the wide variety of existing techniques used in Street skateboarding (Anlauff et al., 2010; Groh et al., 2015).

At present, machine learning is becoming increasingly employed as an efficient solution for problems in several fields of knowledge. In the biomedical sector, artificial neural networks (ANNs) are used in situations involving biometrics (face, iris, fingerprint and voice recognition), image processing (convolutional neural networks), and modeling biological structures with deep neural architecture (Haykin, 2001). The present study used neural networks to develop a classification method for skateboarding trick signals, using artificial acceleration signals as samples. A multilayer feed-forward neural network (MFFNN), allied with a supervised learning algorithm (scaled conjugate gradient, SCG), was used to classify 543 artificially generated signals.

## Methods

### Artificial samples

Artificial signals were generated using the software MATLAB 2015 and Signal Processing Toolbox, based on the findings of previous studies (Groh et al., 2015) and considering what is known about the acceleration and movement typifying each of the tricks chosen to incorporate in this study. Geometric representations of the real experience were modeled through the interpolation of triangles, with base and height determined by a Gaussian distribution (within a limit inferred by inspection of the reference signals). An artificial signal compatible with the references used was obtained through calculating the moving average (MA of 2nd order) of the geometric representations. Allowing the base and height values of the triangles for a random Gaussian distribution permitted the generation of different signals with a similar appearance to the real phenomenon. The method proved to be efficient in generating a signature similar to that found in the literature (Groh et al., 2015). This process was performed for each trick class and its three signatures (X, Y and Z). The artificial signals generated were later added with Gaussian noise to mimic the noise usually captured during the acquisition of physical phenomena. Thus, a total of 543 signals, 181 X-axis, 181 Y-axis, 181 Z-axis, were generated. Figure 1a shows the orientation of the IMU axis of measurement in reference to the fictional skateboard, and Figure 1b shows the type of signatures produced for each axis and class.

Five classes were used for development of the trick classifiers. A total of 181 tricks were used with a random distribution between 50 and 30 samples per class, being 32 Nollie (NOLLIE), 42 Nollie Backside Shove it (NSHOV), 37 Kickflip (FLIP), 32 Backside Shove it (SHOV), and 38 Ollie (OLLIE). The regular reference stance frame (foot positioning) was adopted for the classes OLLIE, NOLLIE and SHOV, and the goofy foot stance for the classes NSHOV and FLIP. This measure was used to test the interference of the individual's stance during the classification process.

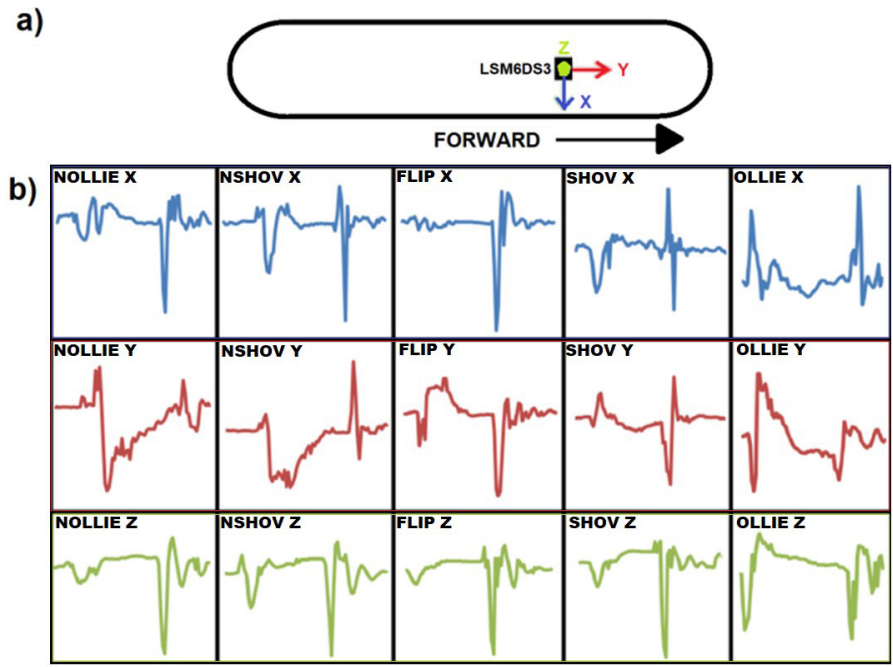

**Figure 1.** (a) Direction of linear acceleration axis; (b) Classes (NOLLIE, NSHOV, FLIP, SHOV, OLLIE) - artificial signatures for acceleration axes (X, Y, Z).

## Detection of events and windowing

The signals were processed using the Signal Processing Toolbox. Windowing was applied using the peaks of acceleration in the artificial signals as reference; perturbations above 5.000 mg (g = 9.8 m/s$^2$; mg = g ×10$^{-3}$) from the baseline correspond to the beginning of an event, while those above 10.000 mg help differentiate the event of interest from other disturbances in the signal.

It was estimated that 90 discrete points of acceleration at most, sampled at 52 samples/s, would be sufficient to window any of the evaluated tricks. These methods and the chosen sample rates were based on similar studies (Anlauff et al., 2010; Groh et al., 2015) and sample rates used in IMU units for human motion detection.

## Classification heuristics

A sample size of 84 artificial signals (28 X-axis signals, 28 Y-axis signals, 28 Z-axis signals) for each of four different Targets (OLLIE, FLIP, NSHOV and SHOV) was used in order to develop a classification method. The NOLLIE class was not used in this step to simplify the problem, given the similarity between OLLIE and NOLLIE (Groh et al., 2015).

The signals were at first divided by axis origin (X, Y or Z) so that patterns in samples from the same class and axis could be extracted. Definition of the samples (Targets) that best represented each class and axis was achieved through cross-correlation (Xcorr) between samples belonging to the same axis and target (FLIP1X, FLIP2X, FLIP3X ... FLIP7X), and the cross-correlation of each possible permutation was calculated for each of the axes of the four classes evaluated.

Cross-correlation between two signals generates a further signal, whereby the maximum peak (σCrr) is a measure of the similarity between the signals cross-correlated. These correlation peaks were used to find the signals that best represented their class (Target). The Targets were composed of those signals that reached higher correlation values among their class and axis, each having the best ranked X, Y and Z signal as a reference for comparison and classification.

It is possible to create a classification method based on which axis best differentiates each class. Figure 2 illustrates the comparison (cross-correlation) between 84 signals and each Target previously assembled. The results are the arithmetic mean of the correlation peaks (σCrr) of each comparison. The highlighted blue line corresponds to the results of the Target when ranking its own class of signals.

These results showed which classes (and their respective axis) are most likely to be falsely classified, also showing how the Z-axis is the most relevant in the classification of any class.

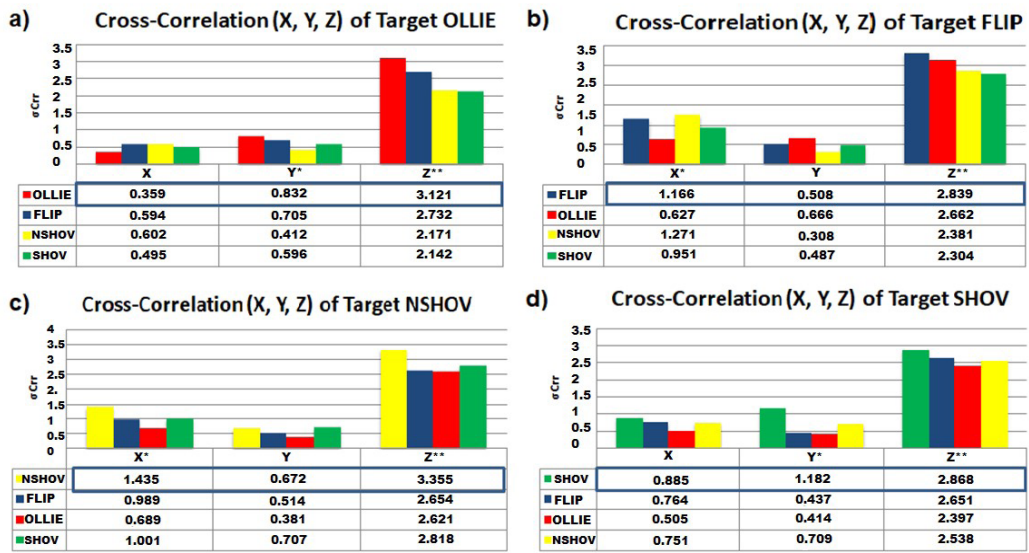

**Figure 2.** Mean values of cross-correlation for all Targets: (a) OLLIE; (b) FLIP; (c) NSHOV; (d) SHOV. **axis with greater significance in the Target classification; *intermediate significance; axis with less significance has no marking; σCrr – peak correlation score.

**Table 1.** Confusion matrix of the correlation coefficients among classes (Z-axis).

| Class | | Correlation coefficients | | | | |
|---|---|---|---|---|---|---|
| 1 | NOLLIE | 0.875 | 0.781 | 0.642 | 0.433 | 0.732 |
| 2 | NSHOV | 0.781 | 0.824 | 0.630 | 0.567 | 0.496 |
| 3 | FLIP | 0.642 | 0.630 | 0.704 | 0.384 | 0.657 |
| 4 | SHOV | 0.433 | 0.567 | 0.384 | 0.633 | 0.164 |
| 5 | OLLIE | 0.732 | 0.496 | 0.657 | 0.164 | 0.861 |
| | | NOLLIE | NSHOV | FLIP | SHIOV | OLLIE |
| | | 1 | 2 | 3 | 4 | 5 |

Green: value of the correlation between the class with itself; Red: class with lower correlation; Blue: classes most likely to be classified incorrectly.

Through these findings a method that uses correlation coefficient values to score the result of each Target in relation to the input signals was implemented. Table 1 shows a confusion matrix where the mean of the correlation coefficients between the signals belonging to the Z-axis are presented. Green depicts the value of the correlation between the class with itself, in red the class with lower correlation, and in blue the classes most likely to be classified incorrectly.

### ANN architecture and training

The adopted heuristics (classification by separate axis) was implemented using the Pattern Recognition Tool of the Neural Network Toolbox. Signals were windowed at 82 points and 459 samples were used for training and validation of the ANN XYZ (Artificial Neural Network trained in the classification of all axes), divided into 367 samples (80%) for training and 92 samples (20%) for validation. ANNs trained with only one of the axes (ANN X, ANN Y and ANN Z) were prepared with 153 samples, 80% for training and 20% for validation. The performance of the trained networks was measured by the minimum cross-entropy values obtained.

The developed ANNs are all MFFNNs that have three layers, an input with 82 neurons, a hidden layer with 28 neurons and a tan-sigmoid transfer function, and an output layer with 5 neurons and softmax transfer function. The learning algorithm used to train the network was SCG (scaled conjugate gradient back propagation) and an illustration of the network can be seen in Figure 3.

The following metric was adopted to define the number of neurons in the hidden layer (Hecht, 1989):

$$N_e \leq \frac{N \times train}{N_{in} + N_{out}} \qquad (1)$$

where $N_e$ is the number of neurons in the hidden layer, N is the number of patterns, $train$ the maximum allowed error (~1%), $N_{in}$ the number of inputs and $N_{out}$ the number of neurons on the output layer.

## Results

Figure 4 shows the decrease in cross-entropy values (low values mean good classification) during training and validation of the ANN Z (a) and the ANN XYZ (b). The validation of ANN XYZ obtained the minimum cross-entropy value in the 42$^{nd}$ iteration, with a value of 0.078475. On the other hand, the performance values were improved in the evaluation of ANN Z; trained on only one axis it obtained a minimum cross-entropy value of 0.019549 in the 23$^{rd}$ iteration. This cross-entropy value represents the best ANNs trained during this study. However, the cross-entropy values of the specialized ANNs were always lower than the non-specialized ANNs.

Figure 5 presents the confusion matrices of both networks, where the total percentage of correctly and incorrectly classified tricks during the training and validation steps can be visualized. The number of true classifications is flagged in green, false positives in red, and circled in yellow are the classes that caused the highest number of false positives for each target. ANN Z (Figure 5a) obtained 98.7% accuracy (error = 1.3%), and the percentage of true classifications of the ANNs X and Y are 94.8% (error = 5.2%) and 96.7% (error = 3.3%), respectively. Figure 5b shows the confusion matrix of ANN XYZ, with true positive rates of 92.8% (error = 7.2%).

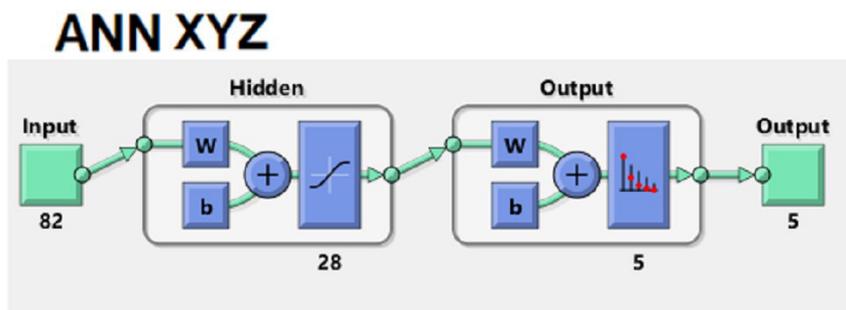

**Figure 3.** Schematic view of the ANNs.

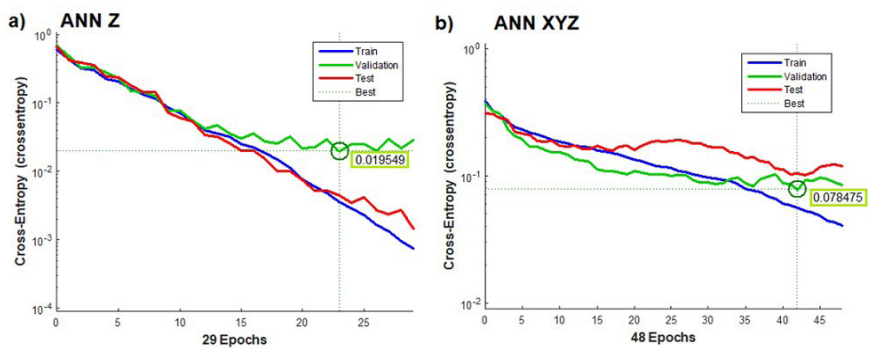

**Figure 4.** (a) Performance of ANN Z - cross-entropy vs. number of iterations (epochs); (b) Performance of ANN XYZ - cross-entropy vs. number of iterations (epochs).

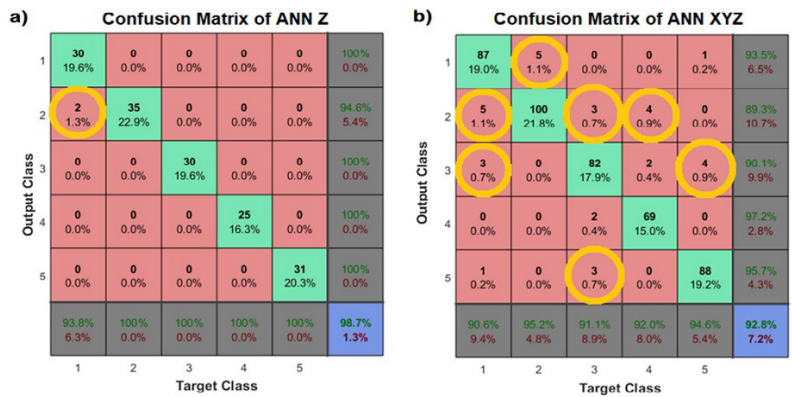

**Figure 5.** (a) ANN Z confusion matrix; (b) ANN XYZ confusion matrix; 1 – NOLLIE; 2 – NSHOV; 3 – FLIP; 4 –SHOV; 5 – OLLIE.

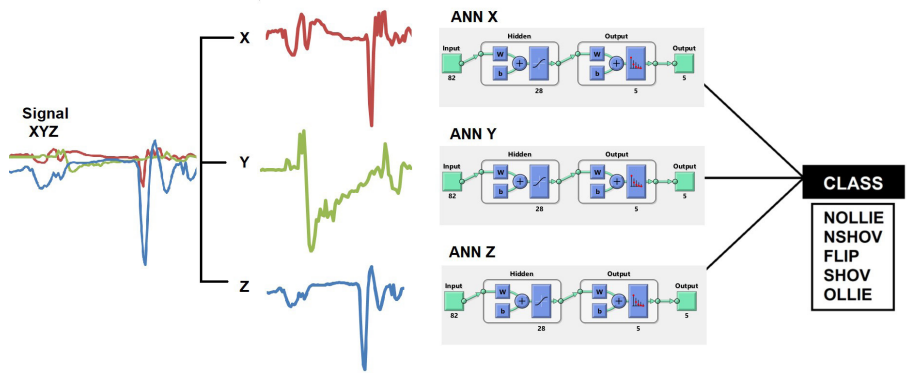

**Figure 6.** Schematic of the classifier formed by specialized ANNs.

**Table 2.** Performance of the classifiers (ANN and Xcorr).

|  | ANN train | ANN validation | Xcorr | Xcorr | Xcorr |
|---|---|---|---|---|---|
| Seconds | 0.1 - 0.8 | 0.001 - 0.003 | 0.004 - 0.006 | 0.045 - 0.050 | 0.068 - 0.070 |
| Nº of samples | 123 | 15 | 3 | 6 | 9 |

The accuracy of the ANN XYZ, when tasked with the classification of a single-axis, was also improved. The classification of the signals corresponding to the X, Y and Z axes obtained the respective results: 88.2% (error = 11.8%), 89.5% (error = 10.5%) and 96.1% (error = 3.9%). Again, discrimination by the Z-axis obtained the best result. However, the classification performance of the X and Y axes is significantly lower for the ANN XYZ than the X and Y ANNs, a result that justifies the choice of using an ANN trained specifically for each axis.

The computational effort and performance of the two classification methods were quantified by measuring the runtime of each classifier (in seconds) as a function of the number of samples. Table 2 presents the performance of the Cross-Correlation (Xcorr) and ANN methods.

Table 2 shows the ANN method classifies with a higher speed (up to $10^{-3}$ seconds) and a greater number of signals than the method using cross-correlation coefficients (σcorr). In addition, performance reduces ($\pm 50 \times 10^{-3}$ seconds) when the classification of more than one trick (3 acceleration signals) is required.

A low error percentage can be obtained with a fast computational response, as classification by ANNs allows the use of only the data coming from a single axis of acceleration (Z-axis). By classifying each of the three axes with the specialized ANNs developed in this study, we can decrease the error percentage to 0.04%, as presented in Figure 6.

## Discussion

The Z-axis was found to be the most effective discriminator among the classes evaluated, given the significantly higher cross-correlation values expressed by the Z-axis in comparison to the other axes, as presented in Figure 2. In addition, analysis of the confusion matrix generated by the correlation coefficients between the created Targets and the input artificial signals (Table 1) revealed a similar pattern to the number of false positives obtained by ANN XYZ.

Neural networks were able to classify 5 tricks (15 signals) in $10^{-3}$ seconds. Although both classifiers (ANN and Xcorr) were able to produce a response in less than $6 \times 10^{-3}$ seconds, the computational effort of the Xcorr classifier increased significantly, $70 \times 10^{-3}$ seconds, as the number of correlations to be calculated increased, making it impossible to develop a fast classifier with a larger class sample space.

The ANN XYZ incorrectly classified 7.2% of the tricks, however, it can be seen from the confusion matrices (Figure 5) that similarities between the acceleration signals for some of the classes affected the classification of the network, such as with the Xcorr method. It is clear that incorrect classifications between classes are due to the presence of similarities, as exposed by the Xcorr method. In the situation of the NOLLIE and NSHOV class, for example, both tricks are executed with a similar foot positioning (Nollie), while for the NSHOV and SHOV class, both tricks have a rotation of 180º around the Z-axis, but in opposite directions due to stance inversion (regular-goofy).

The efficiency of the classifier developed in this study compares favorably with those from previous studies. Groh et al. (2015) achieved 97.8% true positives for the Support Vector Machine and Naive Bayes classifiers, categorizing tricks into 6 different classes, with each trick having 54 features calculated, such as the mean, variance, kurtoses, and spectral density, among other parameters obtained by the IMU signal analysis. A study by Anlauff et al. (2010) obtained 97% true positives using an LDA classifier, categorizing into 3 different classes. During classification seven features were extracted from each event for processing (3D accelerometer, 3D gyroscope, force sensing resistor - FSR). The designed Tilt 'n' Roll classifier proved a good choice for real-time application, however, the classifier was only able to differentiate two classes of tricks using the data from the IMU sensor, with the FSR required to classify the extra third class (Bail). Both studies highlighted the need for a classifier that could cover a greater number of classes without losing computational efficiency (Anlauff et al., 2010; Groh et al., 2015).

The proposed classification by axis proved to be an efficient procedure with minimum dimension use, low-density separation and promising results for real-time application. The combination of IMU and machine learning as a pattern recognition method can perhaps provide real-time activity and logging. However, efficiency can be lost due to overfitting and high density in class separation when working with large data groups with a multilayer perceptron trained in a supervised manner.

The use of deep neural networks (deep learning) and semi-supervised training (Pseudo-Label) is recommended for future studies. Both the machine learning techniques have low density in class separation, low computational cost, and a high classification rate for cases with a large group of classes (Bengio et al., 2012; Donghyun, 2013).

Currently available technologies, whether being the quality of IMU or efficiency of artificial intelligence, can create extremely useful tools, especially in applications involving sensors and human motion recognition. More realistic tools can be developed with high positive rate classification and real time response for CAA in skateboarding and other disciplines.

## Acknowledgements

This research was supported by the Coordination for the Improvement of Higher Level Personnel (CAPES), and the National Council for Scientific and Technological Development (CNPq). The authors would like to thank the researchers from the Microgravity Center at the Pontifical Catholic University of Rio Grande do Sul (PUCRS).